\documentclass[11pt,dvips]{article}

\usepackage{amsmath,epsfig,ulem}
\usepackage{feynarts}
\usepackage{rotating}
\usepackage{cite}
\usepackage{hhline}
\usepackage{array}
\usepackage{amssymb}
\usepackage{color}
\usepackage{dsfont}
\usepackage{equations}
\usepackage{url}

\begin{document}
\tolerance=100000
\setcounter{page}{1}
\interfootnotelinepenalty=10000

\newcommand{\mathbold}[1]{\mbox{\boldmath $\bf#1$}}

\def\tablename{\bf Table}%
\def\figurename{\bf Figure}%
\def\ifmath#1{\relax\ifmmode #1\else $#1$\fi}
\def\ls#1{\ifmath{_{\lower1.5pt\hbox{$\scriptstyle #1$}}}}

\newcommand\phm{\phantom{-}}
\newcommand{\sts}{\scriptstyle}
\newcommand{\ngs}{\!\!\!\!\!\!}
\newcommand{\rb}[2]{\raisebox{#1}[-#1]{#2}}
\newcommand{\CP}{${\cal CP}$~}
\newcommand{\sbomu}{\frac{\sin 2 \beta}{2 \mu}}
\newcommand{\kmol}{\frac{\kappa \mu}{\lambda}}
\newcommand{\s}{\\ \vspace*{-3.5mm}}
\newcommand{\lsim}{\raisebox{-0.13cm}{~\shortstack{$<$\\[-0.07cm] $\sim$}}~}
\newcommand{\gsim}{\raisebox{-0.13cm}{~\shortstack{$>$\\[-0.07cm] $\sim$}}~}
\newcommand{\kr}{\color{red}}

\renewcommand{\Re}{\thinspace{\rm Re\thinspace}}
\renewcommand{\Im}{\thinspace{\rm Im\thinspace}}
\newcommand{\imag}{\Im}
\newcommand{\real}{\Re}
\def\nicefrac#1#2{\hbox{$\frac{#1}{#2}$}}
\def\half{\nicefrac{1}{2}}

\renewcommand{\theequation}{{\rm \thesection.\arabic{equation}}}

\newcommand{\LA}{{\begin{boldmath}$\leftarrow$!! \end{boldmath}}}
\newcommand{\RA}{{\begin{boldmath}!!$\rightarrow$ \end{boldmath}}}

\newcommand{\mycaption}[2][]{{\begin{center} \parbox{17cm}{{\bf \caption[#1]{\it {#2}}}} \end{center} }}

\begin{titlepage}

\begin{flushright} 
{\footnotesize{\it{
DESY 08-030\\
LPT-ORSAY 08-31\\ 
PITHA 08/06 \\
PSI-PR-08-03\\} 
\today}}
\end{flushright}

\vspace{0.6cm}

\begin{center}
{\Large \bf Squark Cascade Decays to Charginos/Neutralinos:  \\[2mm]
            Gluon Radiation}                         \\[1cm] 
{\large R. Horsky$^1$, M. Kr\"amer$^1$, A. M\"uck$^2$, and
        P.M. Zerwas$^{1,3,4}$}\\[1cm]
{\it $^1$ Institut f\"ur Theoretische Physik E, RWTH Aachen, D-52056 Aachen, Germany \\
     $^2$ Paul Scherrer Institut, W\"urenlingen und Villigen, Ch-5232 Villigen PSI, Switzerland\\
     $^3$ Deutsches Elektronen--Synchrotron DESY, D--22603 Hamburg, Germany\\
     $^4$ Laboratoire de Physique Th\'{e}orique, U. Paris-Sud, F-91405 Orsay, France }
\end{center}

\vspace{1.8cm}

\begin{abstract}
\noindent
The momentum spectrum and the polarization of charginos and
neutralinos in squark decays are affected by gluon radiation in the
decay process $\tilde{q} \to q \tilde{\chi} (g)$. We determine these
corrections and study their impact on the [$q\ell$] invariant mass
distributions for leptonic $\tilde\chi$ decays. The higher-order
corrections, though small in general, can be sizeable near pronounced
edges of the final-state distributions.
\end{abstract}

\end{titlepage}

\newpage

\noindent
\section{Introduction}
The properties of supersymmetric particles, colored and non-colored,
can be studied at the LHC in cascade decays of squarks and gluinos,
which have large production cross sections in high-energy hadron
collisions. In particular, squarks that are lighter than gluinos
decay exclusively to quarks and charginos or neutralinos, $\tilde{q}
\to q \tilde{\chi}$.  Subsequent decays of neutralinos into final
states with two leptons are especially well suited to investigate
properties of the supersymmetric spectrum.  Invariant mass
distributions of the final-state leptons and jets can be exploited,
for example, for precision measurements of supersymmetric particle
masses (see {\it e.g.}  \cite{Hinchliffe:1996iu, Allanach:2000kt,
  Gjelsten:2004ki, Weiglein:2004hn}). Moreover, correlations among the
particles in the decay chains allow spin measurements of intermediate
particles~\cite{Barr:2004ze, Smillie:2005ar, Athanasiou:2006ef,
  Wang:2006hk}.

In the present article we concentrate on one aspect in the multitude
of theoretical challenges involved in the description of
supersymmetric cascade decays, {\it i.e.} the super-QCD (SQCD)
radiative corrections at next-to-leading order (NLO) and their impact
on energy distributions, polarization degrees, and on the shape of
quark-lepton invariant mass distributions.  Precise theoretical
predictions for the shape of the distributions are required also at
some distance away from kinematic edges and thresholds as to allow for
sufficiently large, statistically significant event samples. Moreover,
the shapes may be needed to resolve possible ambiguities in mass
determinations from kinematic endpoints~\cite{Gjelsten:2006tg}, they
are crucial for the spin determination~\cite{Barr:2004ze}, and they
allow to test models of supersymmetry breaking \cite{Kitano:2006gv}.
We investigate the impact of QCD radiation on the shape of
distributions in NLO perturbation theory and derive, largely,
analytical results which provide us with the proper understanding of
the SQCD effects.  Extending the Monte Carlo calculation of QCD
effects in Ref.\cite{Miller:2005zp}, we include gluon radiation off
the final-state quarks.  [Hadronic jet fragmentation which must
finally be added is beyond the scope of the present study, but the
effect can be estimated by means of QCD power corrections.]

We shall, for definiteness, focus on cascades of $L$-type squarks,
${\tilde{q}}_L$, of the first two generations without noticeable $L/R$
mixing and consider the decay chain
\begin{equation}\label{eq:cascade}
 {\tilde{q}}_L \to q{\tilde\chi}^0_2 \to q\ell {\tilde\ell}_R \to 
                   q \ell \ell {\tilde\chi}^0_1\,,
\end{equation}
which has been discussed in great detail in the literature for the
SPS1a/a$'$ benchmark points \cite{Allanach:2002nj,
  AguilarSaavedra:2005pw}. The neutralino ${\tilde{\chi}}^0_1$ is
assumed to be the lightest supersymmetric particle (LSP) and thus is
stable in R-parity conserving theories.  Since the Born terms can be
factored out, the NLO analysis applies, without modifications, to all
neutralino and chargino decay chains of the same form. $R$-type
squarks, ${\tilde{q}}_R$, decay in large parts of the supersymmetric
parameter space, particularly at the benchmark points SPS1a/a$'$, to
the LSP, ${\tilde{q}}_R \to q {\tilde\chi}^0_1$, and do not develop
cascades.

Spin correlations are important for the proper description of the
cascade decay (1.1). Since only the gaugino components of the
neutralino ${\tilde{\chi}}^0_2$ couple effectively to the squark/quark
current if quark masses are neglected, the ${\tilde{q}}_L$ squarks
decay to left-handedly polarized $q_L$ quarks.  By angular momentum
conservation also the neutralino ${\tilde\chi}^0_2$ is polarized
left-handedly at the Born level.  Gluon radiation in the decay process
leads to an admixture of right-handed neutralinos. The polarization is
reflected in the energy and angular distributions of the decay
products.

\medskip

The paper is organized as follows. In section~\ref{sec:sqcd} we shall
present the results for the NLO SQCD corrections to the decay
$\tilde{q}_L \to q {\tilde{\chi}}$ and their impact on the $\tilde\chi$
energy distribution and polarization.  The
phenomenology of the complete $\tilde{q}_L$ decay chain (1.1),
including the spin correlations, is discussed in
section~\ref{ssec:qlchain}. In section~\ref{ssec:pp} we analyze
the decay chains in the broader framework of squark production at the
LHC, including $R$-squarks and anti-squarks. We conclude
in section~\ref{sec:conc}.

\begin{boldmath}
\section{Supersymmetric QCD corrections to the decay 
  $\tilde{q} \to q {\tilde{\chi}}$}\label{sec:sqcd}
\setcounter{equation}{0}
Squark\end{boldmath} and gluino decays are affected by SQCD
corrections.  For their mutual decay modes, dominating if
kinematically allowed, the NLO corrections were determined in
Ref.~\cite{Beenakker:1996dw}. The SQCD corrections to the partial
widths for the electroweak squark decay channels,
\begin{equation}
   \tilde{q} \to q + {\tilde{\chi}} + (g) \,,                          
\end{equation}
were calculated in Refs.~\cite{Hikasa:1995bw, Kraml:1996kz,
  Djouadi:1996wt, Beenakker:1996de}: Virtual gluon and gluino/squark
exchanges renormalize the $q \tilde{q} \tilde{\chi}$ vertex and
additional gluons are radiated in the final state, {\it cf.}
Fig.~\ref{Fig:graphs}.

\begin{figure}
\begin{center}

\unitlength=0.7bp

\begin{feynartspicture}(190,154)(1,1)

\FADiagram{}
\FALabel(1.5,13)[t]{$\tilde q$}
\FALabel(14.274,10.)[l]{$\tilde q$}
\FAProp(0.,9.8)(6.5,9.8)(0.,){/Straight}{0}
\FAProp(0.,10.2)(6.5,10.2)(0.,){/Straight}{0}
\FAProp(20.,14.)(13.,14.)(0.,){/Straight}{-1}
\FALabel(21.5,13.5)[b]{$\; q$}
\FAProp(20.,6.)(13.,6.)(0.,){/Straight}{0}
\FAProp(20.,6.)(13.,6.)(0.,){/Sine}{0}
\FALabel(21,7.3)[t]{$\;\;\; \tilde \chi$}
\FAProp(6.4,10.)(12.9,14.)(0.,){/Straight}{0}
\FAProp(6.5,10.)(13.0,14.)(0.,){/Cycles}{0}
\FALabel(9.339,12.9678)[br]{$\tilde g$}
\FAProp(6.5,10.)(13.,6.)(0.,){/Straight}{1}
\FALabel(9.208,6.8192)[tr]{$q$}
\FAProp(13.2,14.)(13.2,6.)(0.,){/Straight}{0}
\FAProp(12.8,14.)(12.8,6.)(0.,){/Straight}{0}
\FAVert(13.,6.){0}
\FAVert(6.5,10.){0}
\FAVert(13.2,14.){0}

\end{feynartspicture}
\begin{feynartspicture}(190,154)(1,1)

\FADiagram{}
\FALabel(1.5,13)[t]{$\tilde q$}
\FAProp(0.,9.8)(6.5,9.8)(0.,){/Straight}{0}
\FAProp(0.,10.2)(6.5,10.2)(0.,){/Straight}{0}
\FAProp(20.,13.993)(13.,14.)(0.,){/Straight}{-1}
\FALabel(21.5,13.5)[b]{$q$}
\FAProp(20.,6.)(13.,6.)(0.,){/Straight}{0}
\FAProp(20.,6.)(13.,6.)(0.,){/Sine}{0}
\FALabel(21,7.3)[t]{\;\; $\tilde \chi$}
\FAProp(6.5,10.)(13.,14.)(0.,){/Cycles}{0}
\FALabel(9.4,13.7769)[br]{$g$}
\FAProp(6.9,10.)(13.4,6.)(0.,){/Straight}{0}
\FAProp(6.1,10.)(12.6,6.)(0.,){/Straight}{0}
\FALabel(9.4,6.8192)[tr]{$\tilde q$}
\FAProp(13.,14.)(13.,6.)(0.,){/Straight}{-1}
\FALabel(14.274,10.)[l]{$q$}
\FAVert(13.,14.){0}
\FAVert(6.5,10.){0}
\FAVert(13.,6.){0}

\end{feynartspicture}

\begin{feynartspicture}(190,154)(1,1)

\FADiagram{}
\FAProp(0.,9.8)(9.,9.8)(0.,){/Straight}{0}
\FAProp(0.,10.2)(9.,10.2)(0.,){/Straight}{0}
\FALabel(1.5,13.)[t]{$\tilde q$}
\FAProp(20.,16.5)(14.618,13.2638)(0.,){/Straight}{-1}
\FALabel(22.3,15.5)[br]{$q$}
\FAProp(20.,3.5)(9.,10.)(0.,){/Straight}{0}
\FAProp(20.,3.5)(9.,10.)(0.,){/Sine}{0}
\FALabel(22.5,5)[tr]{\;\;\; $\tilde \chi$}
\FAProp(20.,10.)(14.618,13.2638)(0.,){/Cycles}{0}
\FALabel(22.3,10.5)[tr]{$g$}
\FAProp(9.,10.)(14.618,13.2638)(0.,){/Straight}{1}
\FALabel(11.4312,12.6286)[br]{$q$}
\FAVert(9.,10.){0}
\FAVert(14.618,13.2638){0}

\end{feynartspicture}
\begin{feynartspicture}(190,154)(1,1)

\FADiagram{}
\FAProp(0.,9.8)(5.4491,9.7904)(0.,){/Straight}{0}
\FAProp(0.,10.2)(5.4491,10.2095)(0.,){/Straight}{0}
\FALabel(1.5,13.)[t]{$\tilde q$}
\FAProp(20,16.5)(12.0059,10.0299)(0.,){/Straight}{-1}
\FALabel(22.,15.5)[br]{$q$}
\FAProp(20.,3.5)(12.0059,10.0299)(0.,){/Straight}{0}
\FAProp(20.,3.5)(12.0059,10.0299)(0.,){/Sine}{0}
\FALabel(20.5,3.)[bl]{$\; \tilde \chi$}
\FAProp(12.5,3.5)(5.4491,9.97)(0.,){/Cycles}{0}
\FALabel(9.5,4.8)[tr]{$g$}
\FAProp(5.4491,9.7904)(12.0059,9.7904)(0.,){/Straight}{0}
\FAProp(5.4491,10.2095)(12.0059,10.2095)(0.,){/Straight}{0}
\FALabel(8.7526,11.)[b]{$\tilde q$}
\FAVert(12.0059,10.0299){0}
\FAVert(5.4491,10.){0}
\end{feynartspicture}

\end{center}
\mycaption{\label{Fig:graphs} Generic Feynman diagrams for virtual and
  real super-QCD corrections to the decay ${\tilde{q}} \to q
  {\tilde{\chi}}$.}
\end{figure}
Vertex corrections neither change the $\tilde{\chi}$ momentum nor its
polarization state, but both are affected by gluon radiation. The
$\tilde{\chi}$ line spectrum of the two-body decay becomes a
continuous spectrum (section~\ref{ssec:dalitz}).  Since in radiative
decays the $\tilde{\chi}$ is not back-to-back anymore with the
polarized zero-mass quark $q$, spin-flipped $\tilde{\chi}$ states are
admixed (section~\ref{ssec:chipol}). While the effect on the
polarization is small, the overall impact of gluon radiation on the
final-state distributions, particularly near edges, is quite
significant, {\it nota bene} in view of the envisaged high-precision
analyses of these cascade modes at the LHC.

\begin{boldmath}
\subsection{$\tilde{\chi}$ Energy
  distribution}\label{ssec:dalitz} 
The \end{boldmath} Dalitz-plot distribution for
the unpolarized decay $\tilde{q} \to q {\tilde{\chi}} g$, normalized
to the Born-level decay width ${\Gamma}_{\rm{B}}$,
\begin{eqnarray}\label{eq:dalitz} 
  \frac{1}{{\Gamma}_{\rm{B}}} \frac{d\Gamma}{dx_q dx_g} =
  \frac{4}{3} \frac{\alpha_{\mathrm{s}}}{\pi} \frac{1}{(1-\kappa)^2}
   & & \left[
  \frac{2 - 2 \kappa (1 + x_q) - x_g (1 + \kappa + x_q)}
       {2 x_g (1 + \kappa - x_{\tilde\chi})} -
  \frac{(1 - \kappa) - x_g}{x_g^2}  
  \right]\,,
\end{eqnarray}
is expressed by $x_{q,g,\tilde{\chi}} = 2
E_{q,g,\tilde{\chi}}/M_{\tilde{q}}$, where $M_{\tilde{q}}$ is the
squark mass and $E_{q,g,\tilde{\chi}}$ are the energies of the quark,
the gluon, and the $\tilde\chi$ in the squark rest frame with
$x_{\tilde\chi}+x_q+x_g=2$ and $2 \sqrt{\kappa} \leq x_{\tilde{\chi }}
\leq 1+\kappa$.  A useful abbreviation is the mass ratio
$\kappa=M^2_{\tilde{\chi}}/M^2_{\tilde{q}}$.  The partial width in
Born approximation, Refs.~\cite{Kraml:1996kz,Djouadi:1996wt},
\begin{equation}
 \Gamma_B = \frac{\alpha}{4} M_{\tilde{q}} (1-\kappa)^2 f^2_{{\tilde{q}} {\tilde{\chi}}} \,,
\end{equation}
depends on the chargino and neutralino mixing parameters
\cite{Choi:2000ta, Choi:2001ww} in the coefficients $f_{{\tilde{q}}
  {\tilde{\chi}}}$.  Gluon radiation collinear with the final-state
quark is described by the first term in Eq.~(\ref{eq:dalitz}) since $1
+ \kappa - x_{\tilde\chi}=M^2_{qg}/M^2_{\tilde{q}}$, where $M_{qg}$ is
the $[qg]$ invariant mass.  The second term damps the infrared
singularity.

Gluon radiation reduces the energy $x_{\tilde{\chi}}$ of the
$\tilde\chi$ according to the distribution
\begin{equation}
   \frac{1}{{\Gamma}_{\rm{B}}} \frac{d\Gamma}{dx_{\tilde{\chi}}} =
\frac{4}{3} \frac{\alpha_{\mathrm{s}}}{\pi} \frac{1}{(1-\kappa)^2} \frac{1}{1+\kappa-x_{\tilde{\chi }}}
\left[
      (1-\kappa) \left(2-x_{\tilde{\chi}}
      \right) 
      \log \left(\frac{2 -x_{\tilde{\chi}} + p_{\tilde{\chi }}}{2 -x_{\tilde{\chi }} -p_{\tilde{\chi }}}\right) 
      + \frac{1}{4} p_{\tilde{\chi }} \left(x_{\tilde{\chi }} -8 + 6\kappa \right)
\right]\,,
\end{equation}
where $p_{\tilde{\chi }}=\sqrt{x_{\tilde{\chi }}^2- 4 \kappa}$ denotes
the scaled neutralino momentum in the squark rest frame.

Collinear and infrared gluon emission generate singularities at the
maximum $\tilde\chi$ energy $x_{\tilde{\chi}} \to
x_{\tilde{\chi}}^{\rm max} = 1+\kappa$.  The leading contribution in
$\delta = x_{\tilde{\chi}}^{\rm max} -x_{\tilde{\chi }}$ can be
resummed by applying techniques analogous to thrust distributions
\cite{Schierholz:1979bt, Binetruy:1980hd, Catani:1991kz} or
lepto-quark decays~\cite{Plehn:1997az}:
\begin{eqnarray}\label{eq:resum}
    \frac{1}{{\Gamma}_{\rm{B}}} \frac{d\Gamma}{dx_{\tilde{\chi}}} &\simeq&
      \frac{4}{3} \frac{\alpha_{\rm s}}{\pi} \frac{1}{\delta} \left[\log{\frac{1}{\delta}} + c \right]\nonumber \\
                                                                  &\Rightarrow&
      \frac{4}{3} \frac{\alpha_{\rm s}}{\pi} \frac{1}{\delta} \left[\log{\frac{1}{\delta}} + c \right]
                  \exp{\left[-\frac{4}{3}\frac{\alpha_{\rm s}}{\pi}
                       \left(\frac{1}{2} {\log}^2 \frac{1}{\delta} + c \log \frac{1}{\delta} \right) \right]}
\end{eqnarray}
with the sub-leading coefficient
\begin{equation}
  c = - \frac{7}{4} + \frac{\log [(1-\kappa)^2]}{(1-\kappa)^2} \,.
\end{equation}
Multiple gluon emission bends the distribution to zero at the
kinematical boundary, corresponding to the Sudakov suppression of a
final state without any gluons. Depending in detail on the effective
value of $\alpha_{\rm s}$, the turn-over point is very close to the
maximum of $x_{\tilde\chi}$.

The $\tilde\chi$ energy distribution is displayed in
Fig.~\ref{Fig:pol}(a) for $L$-squark decays to ${\tilde{\chi}}^0_2$,
with parameters adopted from SPS1a$'$ ({\it i.e.} $\kappa=0.108$) and
the QCD coupling set to $\alpha_\mathrm{s} (M_{\tilde{q}})= 0.093$.

\begin{boldmath}
\subsection{$\tilde\chi$ Polarization vector}\label{ssec:chipol}
Gluon \end{boldmath} 
radiation at NLO also affects the $\tilde{\chi}$ polarization.
At the Born level the $\tilde{\chi}$ polarization in $\tilde{q}_L$
decays must be left-handed.  While the $\tilde{\chi}$ spin does not
flip for collinear and infrared gluon emission, because the
intermediate left-chiral quark is effectively on-shell, the
$\tilde{\chi}$ spin can flip for non-collinear/non-infrared
final-state configurations, with the angular momentum balanced by the
emitted hard gluon.

The polarization vector $\vec{\mathcal{P}}$$[\tilde{\chi}]$ is defined in
the ${\tilde{\chi}}$ rest frame.  The $x$-axis is chosen as the
$\tilde\chi$ flight direction in the $\tilde{q}$ rest frame, 
the [$x,y$]-plane is identified with the decay plane, 
the $y$-axis pointing to the half-plane of shortest
rotation from the $\tilde\chi$ to the $q$ momentum vector. The
polarization vector lies within the decay plane, with components
parallel, $\mathcal{P}_\parallel=\mathcal{P}_x$, and transverse,
$\mathcal{P}_\perp=\mathcal{P}_y$, to the $\tilde\chi$ flight
direction while the perpendicular polarization component vanishes
($\mathcal{P}_z = 0$).

The two components of the polarization vector for fixed
$x_{\tilde\chi}$ can be cast into the form
\begin{eqnarray} \nonumber
   \mathcal{P}_\parallel (x_{\tilde\chi}) = 
\frac{1}{4}
     \frac{1}{(1-\kappa)^2} \frac{1}{\left(1+\kappa -x_{\tilde{\chi}}\right) p_{\tilde{\chi }}} 
 &&  \, \Bigl[ 
               4 \left(2-x_{\tilde{\chi }}\right) 
            \left(\kappa \left[4-x_{\tilde{\chi }}\right] - x_{\tilde{\chi }}\right) 
            \log \left(\frac{2 -x_{\tilde{\chi}} + p_{\tilde{\chi }}}{2 -x_{\tilde{\chi }} 
                             -p_{\tilde{\chi }}}\right) +
             \Bigl. \\ \label{eq:P_of_x_chi} && \, \, \, \Bigl.
            \left(
                  \left[8-x_{\tilde{\chi }}\right] 
                   x_{\tilde{\chi}}- 4 \kappa  \left[7 - 2 x_{\tilde{\chi }} \right]
            \right) 
            p_{\tilde{\chi }}
      \Bigr] / \mathcal{N}      \,,
\\ \nonumber
   \mathcal{P}_\perp (x_{\tilde\chi}) = 
\frac{\pi}{8} \frac{1}{(1-\kappa)^2} \frac{\sqrt{\kappa }}
     {\sqrt{1+\kappa -x_{\tilde{\chi }}} \, \, p_{\tilde{\chi }}} \, 
 && 
      \left[ 32 +
            20 \kappa 
               -16 \sqrt{1+\kappa -x_{\tilde{\chi}}} 
            \left(2-x_{\tilde{\chi }}\right) -      x_{\tilde{\chi }} 
                                                              \left(32 -3 x_{\tilde{\chi }}\right)
     \right] / \mathcal{N}  \, .
\end{eqnarray}
The normalization factor $\mathcal{N}$ denotes the distribution
${(\frac{4}{3} \frac{\alpha_{\mathrm{s}}}{\pi} \, \Gamma}_B)^{-1}
d\Gamma/dx_{\tilde\chi}$ at the point $x_{\tilde\chi}$.  For fixed
$x_{\tilde\chi} < x_{\tilde\chi}^{\rm max}$ the polarization vector is
independent of $\alpha_{\mathrm{s}}$ to leading order and depends only
on the radiation dynamics.  In the infrared and collinear limits of
gluon radiation, $x_{\tilde\chi} \to x_{\tilde\chi}^{\rm max}$, the
Born values are approached again,
\begin{equation}
\left.
\begin{array}{cclccc}
   \mathcal{P}_\parallel &\to& 
   - 1\; + \!\!& \begin{displaystyle}
           \frac{\kappa}{2 (1 - \kappa)^2} 
   \frac{\delta}{ \log(1/\delta)}
   \end{displaystyle}
   & \to & -1 \\[4mm]   
   \mathcal{P}_\perp     &\to &   
         &    
         \begin{displaystyle}
         \frac{3 \pi}{8} \frac{\sqrt{\kappa}}{(1 - \kappa)} 
   \frac{\sqrt{\delta}}{ \log(1/\delta)}
   \end{displaystyle}
   & \to & 0   
\end{array}
\right\}  
\quad {\rm for}\quad  x_{\tilde\chi} \to x_{\tilde\chi}^{\rm max} =1+\kappa\,.
\end{equation}
In the opposite limit, $x_{\tilde{\chi}} \to x_{\tilde{\chi}}^{\rm
  min} = 2 \sqrt{\kappa}$, the polarization components approach the
values $\mathcal{P}_\parallel \to 0$ and $\mathcal{P}_\perp \to
-\pi/4$. The two components of the polarization vector are displayed
in Fig.~\ref{Fig:pol}(b), again for the mass pattern of the SPS1a$'$
scenario ($\kappa=0.108$).
\begin{figure}
\begin{center}
\includegraphics[width=17cm]{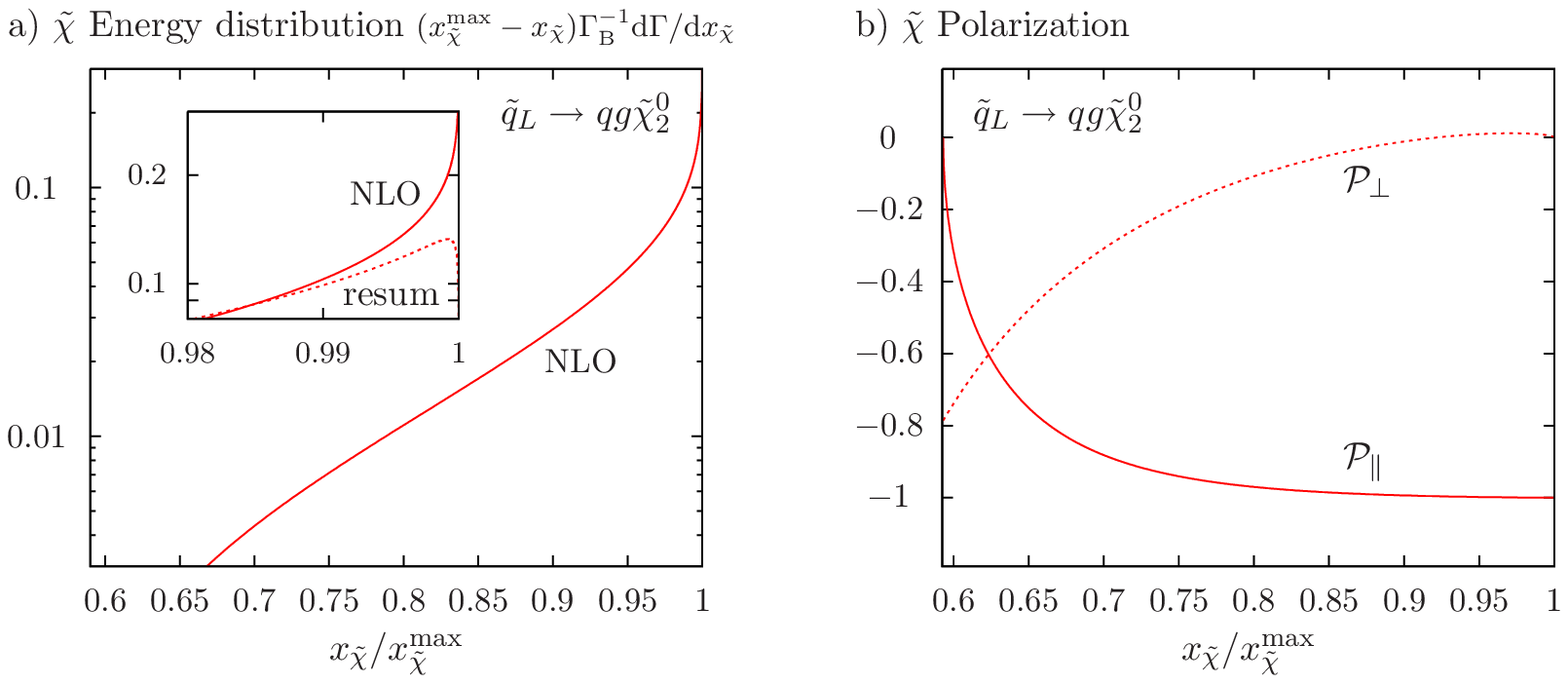}
\end{center}
\mycaption{\label{Fig:pol} (a) The differential decay rate for
  ${\tilde{q}}_L \to q g {\tilde\chi}^0_2$ as a function of the
  ${\tilde\chi}^0_2$ energy $x_{\tilde\chi}$ at the benchmark point
  SPS1a$'$ ($M_{\tilde{\chi}}/M_{\tilde{q}}=0.329$).  The curve
  labeled ``resum'' shows the damping near $x_{\tilde\chi}^{\rm max}$
  from multiple gluon radiation according to Eq.~(\ref{eq:resum}).
  (b) The longitudinal and transverse polarization of the neutralino
  ${\tilde\chi}^0_2$.}
\end{figure}
\\[-2mm]

\begin{boldmath}
  \subsection{Average $\tilde\chi$ energy and polarization}
The\end{boldmath} values of the average energy and the polarization
components of $\tilde\chi$ for the inclusive NLO prediction are
obtained by integrating over the Dalitz plot and including the 2-body
decay limit.  Both, the QCD corrected 2-body and the 3-body part
of the calculation are infrared divergent.  Analogously to the methods
employed in Refs.~\cite{Kraml:1996kz, Djouadi:1996wt}, small quark and
gluon masses are introduced to regularize the singularities.  It is
sufficient to know the regularized form of the different contributions
for the unpolarized case. Since the polarization effects are not
infrared-sensitive, the only divergent integrals are those already
encountered in the unpolarized results.  The averaged observables
$\langle x_{\tilde{\chi}} \rangle$, $\langle \mathcal{P}_\parallel
\rangle$, and $\langle \mathcal{P}_\perp \rangle$ globally reflect the
deviation from the Born prediction to order ${\alpha}_\mathrm{s}$, and
they are given by
\begin{eqnarray}
   \langle x_{\tilde\chi} \rangle = x^{\rm max}_{\tilde\chi} -
   \frac{1}{9} \frac{\alpha_{\mathrm{s}}}{\pi} \frac{1}{(1-\kappa)^2 } 
   \left[(1-\kappa) (4-\kappa  [17-7 \kappa]) - 6 (2-\kappa) \kappa ^2 \log \kappa\, \right]
\end{eqnarray}
and
\begin{eqnarray} \nonumber
   \langle \mathcal{P}_\parallel \rangle = -1 
-\frac{1}{9} \frac{\alpha_{\mathrm{s}}}{\pi} \frac{1}{(1-\kappa)^2}
&& \, 
\left[ 4 \pi ^2 (\kappa -3) \kappa -48 (\kappa -3) \kappa \text{Li}_2\left(\sqrt{\kappa }\right) 
      +12 (1-\kappa  [8- 3 \kappa]) \text{Li}_2(\kappa )
      \right. \\ && \left. \, \,  \nonumber   
     -3 \sqrt{\kappa }\left(22[1+\kappa] - \sqrt{\kappa }[45 -{\kappa}]\right)
       +6 (1- \kappa) (11 - 3 \kappa )
     \log \left(\sqrt{\kappa }+1\right)
      \right. \\ && \left. \, \,  
     +3 \left(4 [1-\kappa]^2 \log (1-\kappa )+ \kappa [4-3 \kappa ] \right)
     \log \,\kappa
\right] \,, \\ \nonumber
   \langle \mathcal{P}_\perp     \rangle =   
\frac{ 2 \alpha_{\mathrm{s}}}{3} \frac{\sqrt{\kappa}}{(1-\kappa)^2}
&& 
\left[ 4+4 \log \,\kappa -4 \kappa
      -\left(1+\sqrt{\kappa }\right) (15-\kappa ) \, 
      \mathrm{E}
      -2 \left(\kappa -7 \sqrt{\kappa }-8\right) 
      \mathrm{K}
\right] \,.
\end{eqnarray}
The function $\text{Li}_2$ is the usual dilogarithm while K and E
abbreviate the complete elliptic integrals of the first and second
kind\footnote{The integrals are defined as $K(z)=\int_0^{\pi/2} (1-z
  \sin^2 x)^{-1/2} dx $ and $E(z)=\int_0^{\pi/2} (1-z \sin^2 x)^{1/2}
  dx $.}, respectively, evaluated for the argument
$\left[\left(1-\sqrt{\kappa}\right)/\left(1+\sqrt{\kappa}\right)\right]^2$.

For the reference point SPS1a$'$ ($\kappa=0.108$) adopted in the
numerical analyses of the next section, the mass-dependent
coefficients of the ${\cal O}(\alpha_{\mathrm{s}})$ contributions take
values
\begin{equation} 
   \langle x_{\tilde\chi} \rangle = x^{\rm max}_{\tilde\chi} - 0.32 \, \frac{\alpha_{\mathrm{s}} }{ \pi}\,,
\end{equation}              
and 
\begin{equation} 
   \langle \mathcal{P}_\parallel \rangle = -1 + 0.02 \, \frac{
     \alpha_{\mathrm{s}} }{ \pi}\,,  \qquad
   \langle \mathcal{P}_\perp     \rangle = + 0.08 \, \frac{ \alpha_{\mathrm{s}} }{3}   \,.
\end{equation} 
The NLO SQCD effect on the average longitudinal component of the
$\tilde\chi$ polarization vector, $\langle \mathcal{P}_\parallel
\rangle$, is very small. This follows from the fact that in the
collinear and infrared regions, in which the density of the Dalitz
plot is maximal, the deviation of the polarization vector from the
Born approximation approaches zero. The average perpendicular
polarization $\langle \mathcal{P}_\perp \rangle$ is positive,
dominated by the contributions from the region of large $\tilde\chi$
energies.

\section{Phenomenological results}
\setcounter{equation}{0}

\begin{boldmath}
\subsection{The ${\tilde{q}}_L$ decay chain}\label{ssec:qlchain}
For \end{boldmath} parameter points like SPS1a/1a$'$, the decays
to lepton/slepton pairs of the first two generations, 
\begin{equation}
   {\tilde\chi}^0_2 \to \ell^+_R + {\tilde{\ell}}_R^- \;\;\;{\rm and}\;\;\;
                        \ell^-_R + {\tilde{\ell}}_R^+ \,,
\end{equation}
are ideal spin analyzers [$R$ denoting right chirality] since the
decays into the $L$-type state ${\tilde{\ell}}_L^\pm$ are
kinematically forbidden, and $L/R$ slepton mixing as well as the
$\tilde{\chi}^0_2$ higgsino component are suppressed. [Small
corrections due to non-zero lepton masses can easily be included.]

For the sake of clarity, we first focus on these $\ell_R^+
{\tilde\ell}_R^-$ decay modes of the neutralino. Since the positron
$\ell^+_R$ is left-handedly polarized, it will be emitted
preferentially anti-parallel to the $\tilde\chi$ spin direction,
\begin{equation}\label{eq:pol_effect}
   \frac{1}{\Gamma} \, \frac{d \, \Gamma}{d \, \cos \theta_s} =
           \frac{1}{2} \, (1 - {\mathcal{P}}[\tilde\chi] \cos\theta_s)\,,
\end{equation}
where $\theta_s$ is the angle between the $\tilde\chi$ spin vector and
the $\ell^+$ 3-momentum, and ${\mathcal{P}}[\tilde\chi]$ the
$\tilde\chi$ degree of polarization. The angular correlation
Eq.~(\ref{eq:pol_effect}) gives rise to an increase of the invariant
mass of the [$q \ell^+$] pair in the decay chain
\begin{equation}
   \tilde{q}_L \to q + {\tilde\chi}^0_2 + (g)
              \to q + \ell^+ + \tilde{\ell}^-_R + (g) 
\end{equation}
compared with the isotropic distribution.  [In this cascade the lepton
$\ell$ is generally termed ``near lepton'' as opposed to the ``far
lepton'' emitted in the subsequent slepton decay $\tilde{\ell} \to 
\ell {\tilde\chi}^0_1$, which will be included later.]

We have implemented the NLO results presented above in a flexible
parton level Monte-Carlo program to calculate arbitrary differential
distributions. Both phase-space slicing and subtraction methods have
been employed to allow for internal cross checks.

In the phase-space slicing approach, approximate real radiation
matrix elements are integrated analytically in the soft/collinear
regions using mass regularization.  After adding the mass-regularized
virtual corrections, the result is finite and the regulating quark and
gluon masses can be set to zero. The hard gluon region is integrated
by standard Monte-Carlo techniques.

Alternatively, a subtraction method has been applied which is
particularly suited to construct NLO parton-level Monte-Carlo
programs.  In the present simple case the entire 3-particle
differential decay width can be used as subtraction term.  For the
observables $\mathcal{O}$ which we consider, the expression in NLO can
be cast in the form
\begin{equation}
\label{eq:subtraction}
 {\langle {\mathcal{O}} \rangle}_{\rm{NLO}} =
  \int_3 d{{\hat{\Gamma}}}^R_3 [{\mathcal{O}}_3 - {\mathcal{O}}_2] +
  \int_2 [d{{\hat{\Gamma}}}_2^V + \int_1 d{{\hat{\Gamma}}}_3^R ] {\mathcal{O}}_2 \,.
\end{equation} 
Here, $d{{\hat{\Gamma}}}_3^R$ is the differential 3-parton decay
width, $d{{\hat{\Gamma}}}_2^V$ the vertex-corrected 2-parton decay
width, both normalized to the NLO width $\Gamma$. Moreover,
${\mathcal{O}}_n$ denotes the observable ${\mathcal{O}}$ calculated
for either $n=3$ or $n=2$ parton final states.  The projection from
the $n+1$ onto the $n$ particle phase space is straightforward: at any
given 3-body phase-space point, ${\mathcal{O}}_2$ is evaluated for a
neutralino momentum which shares the direction of flight with the
neutralino momentum of the 3-body phase-space point. The simple 2-body
kinematics fixes the neutralino and quark momenta completely so that
${\mathcal{O}}_2$ can be calculated unambiguously.  The first integral
in Eq.~(\ref{eq:subtraction}) is finite by construction and can be
integrated by Monte-Carlo techniques. The second and the third
integral can be evaluated analytically [in mass-regulated form to
isolate the infrared/collinear divergences, which cancel in the sum of
the virtual contribution and the integrated subtraction term]. Since
polarization effects are not infrared sensitive, it is sufficient to
use the unpolarized differential decay width $d{{\hat{\Gamma}}}_3^R$
as a subtraction term even for observables including
spin correlations. In this case, ${\mathcal{O}}_2$ is evaluated with
the tree-level polarization vector, which provides the correct limit
of the 3-body kinematics in the soft/collinear regions. The results
derived by using the subtraction method agree with those obtained from
phase-space slicing.

\begin{figure}
\begin{center}
\includegraphics[width=17cm]{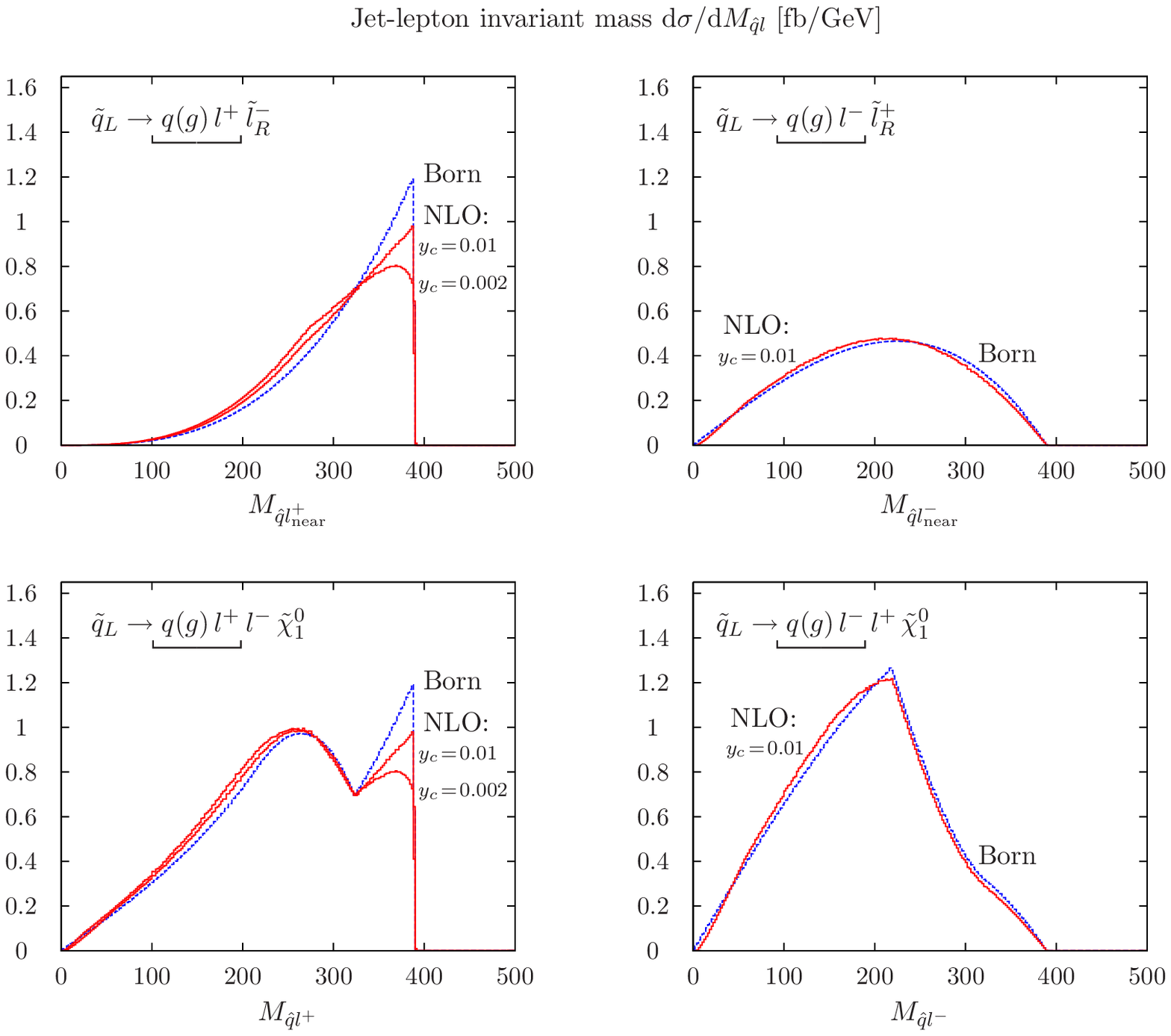}
\end{center}
\vspace{-0.75cm} \mycaption{\label{Fig:inv_mass} The
  $[{\hat{q}}l^\pm]$ invariant mass distributions for positively (left
  panels) and negatively (right panels) charged leptons is shown at
  SPS1a$'$ [specifically for $\tilde{u}_L$ masses and first and second
  generation leptons with vanishing masses]. The upper panels include
  only ``near'' leptons while the lower panels display the sum of
  ``near'' and ``far'' leptons. The NLO predictions are represented by
  the solid lines for $y$-cut parameters $y_{\rm c}$ = 0.01 and 0.002,
  the Born result [approximately equivalent to NLO with $y_{\rm c}
  \gsim 0.05$] by the dashed line.}
\end{figure}

When gluon radiation is switched on, the [$q \ell^+$] invariant mass
distribution is softened. To quantify this effect, the $q$-jet must be
defined properly in an infrared-safe way.  We will combine the $q$ and
$g$ partons in the collinear and infrared regimes to a single
$\hat{q}$-jet if their scaled invariant mass $y = M^2[\hat{q}=qg] /
M^2_{\tilde{q}} $ is less than $y_\mathrm{c}=0.01$. For events with
non-collinear hard gluon emission we identify $\hat{q}$ with the
leading parton jet, {\it i.e.} the jet with the highest energy in the
squark rest frame, either $q$ or, rarely, $g$.

The invariant mass distributions $M^2[\hat{q} \ell^\pm]$ are displayed
as solid curves in Fig.~\ref{Fig:inv_mass}.  In the upper left panel
the distribution of the positively charged ``near'' leptons, produced
in ${\tilde{\chi}}^0_2$ decays, is shown. The dashed curve represents
the invariant mass distribution at the Born level for comparison. The
peak of the distribution at high invariant mass is more pronounced
than the triangular-shaped tree-level distribution with
spin correlations ignored~\cite{Richardson:2001df}.  As expected,
the NLO corrections lead to the transfer of events to smaller
invariant masses compared to the tree-level prediction.

The transition from the 1-jet to the 2-jet class of events gives rise
to the tiny kink in the $[\hat{q} \ell^+_{\rm near}]$ distribution in
Fig.~\ref{Fig:inv_mass}. Varying the experimentally defined
jet-resolution parameter $y_\mathrm{c}$ reproduces an\-ti\-ci\-pa\-ted
modifications of the mass distributions: the distribution is softened
for smaller $y_\mathrm{c}$, and the peak rounded off, exemplified in
Fig.~\ref{Fig:inv_mass} for $y_\mathrm{c}= 0.002$, as a larger
fraction of events with respect to the default choice
$y_\mathrm{c}=0.01$ moves from the 1-jet to the 2-jet class. For
larger $y_\mathrm{c}$ the distribution is sharpened, moving back
towards the Born distribution, as more events are attributed to the
1-jet class. The NLO distribution becomes almost identical to the LO
prediction for $y_\mathrm{c} \gsim 0.05$.  Quite similar results for
the invariant mass distributions are found when using a cone
algorithm~\cite{Sterman:1977wj} defined in the squark rest frame.

Parallel to the $\tilde\chi$ energy distribution, exponential Sudakov
damping of the invariant [$\hat{q} \ell^+$] mass distribution is
expected in a margin of one percent below the pronounced kinematical
edge when multiple gluon radiation is switched on. Likewise, hadronic
jet fragmentation will give rise to smearing effects, generally
accounted for, see {\it e.g.} Ref.~\cite{Heister:2003aj}, by shifting
the perturbative prediction of the distribution downwards by a
relative amount of approximately $1\,{\rm GeV}\,/M_{\tilde{q}}$.  For
squarks decaying through electroweak channels, as at the present
reference point, non-zero width effects $\propto \Gamma_{\tilde{q}} /
M_{\tilde{q}}$ are also limited to about one percent, on both sides of
the kinematical edge, though.\footnote{The accuracy of the
  narrow-width approximation for cascade decays has been addressed in
  Ref.~\cite{Kauer:2007nt}.}  These effects will round off the peak of
the distribution at the edge for $y_\mathrm{c} \gsim 0.01$, with the
overall impact limited however to a margin at the one-percent level
for squark masses in excess of 500 GeV. With details depending on the
jet definition, the modification will qualitatively be similar to the
thrust distribution in jet analyses of $e^+ e^-$ annihilation at high
energies \cite{Heister:2003aj}.  The explicit Monte Carlo simulation
of QCD effects in Ref.~\cite{Miller:2005zp} includes only gluon
radiation off the squarks which turns out, as expected, to have a very
small effect on the shape of the distributions.

The experimentally observed lepton $\ell^+$ may also be the decay
product of the slepton in the chain\footnote{Electroweak radiative
  corrections to the decay $\tilde{\chi}_2^0\to l^+ l^-
  \tilde{\chi}_1^0$ have been presented in Ref.~\cite{Drees:2006um}.}
${\tilde\chi}^0_2 \to \ell^-_R \, {\tilde\ell}^+_R \to \ell^-_R \,
\ell^+_R \, {\tilde\chi}^0_1$.  For this ``far'' lepton $\ell^+$, the
daughter of a spinless parent, the spin characteristics in the
distributions are largely washed out and only the average energy
reduction of the slepton due to spin correlations in the primary
neutralino decay is effective.

Adding up the contributions from ``near'' and ``far'' $\ell^+$
leptons, which cannot be separated experimentally on an event-by-event
basis, we arrive at the distributions displayed in the lower left
panel of Fig.~\ref{Fig:inv_mass}. Due to the rather small mass
difference between the selectron/smuon and the LSP
${\tilde{\chi}}_1^0$ in SPS1a$'$, the lepton from the slepton decay is
relatively low-energetic.  Thus the ``far'' lepton adds to the
distribution only at low invariant masses and the signature of the
``near'' lepton is still significant at high invariant masses.

If, instead, negatively charged leptons are observed the shapes of the
two corresponding distributions are rather different. Since the
negatively charged leptons are right-handedly polarized the sign in
Eq.~(\ref{eq:pol_effect}) flips and the ``near'' $\ell^-$ is
preferentially emitted parallel to the $q$-jet, in contrast to the
anti-parallel emission of positively charged leptons. The energy of
the negatively charged lepton $\ell^-$ is therefore, on average,
reduced in the squark rest frame for polarized ${\tilde\chi}^0_2$
decays. For both reasons, the $[\hat{q} \ell^-]$ invariant mass is
reduced on average compared to the prediction without spin
correlations~\cite{Richardson:2001df}. The reduction is slightly
enhanced by gluon emission as demonstrated in the upper right panel of
Fig.~\ref{Fig:inv_mass}.  The distribution for the sum of ``near'' and
``far'' $\ell^-$ leptons is shown in the lower right panel.  Its shape
can easily be distinguished from the $\ell^+$ results at large
invariant masses.

Spin correlations are clearly important for the description of the
invariant mass distributions; without taking spin correlations into
account, all panels in Fig.~\ref{Fig:inv_mass} would be identical.
\begin{figure}
\begin{center}
\includegraphics[width=8.25cm]{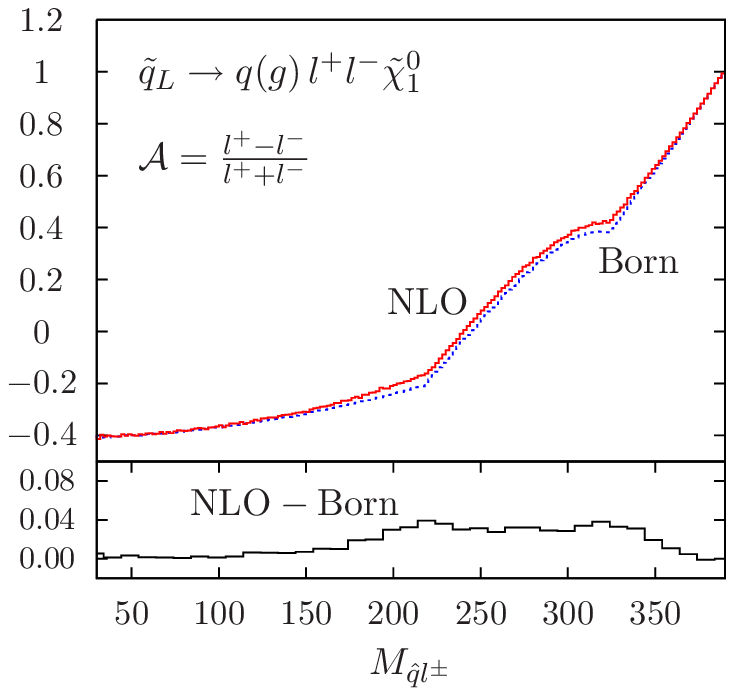}
\end{center}
\mycaption{\label{Fig:asym} The asymmetry Eq.~(\ref{eq:asym}) as a
  function of the $[\hat{q} l^\pm]$ invariant mass at SPS1a$'$ for
  $\tilde{q}_L$ decays [specifically for $\tilde{u}_L$ masses] into
  first and second generation lepton pairs (no anti-squark decays
  included).}
\end{figure}
Comparing the final distributions in the lower panels an asymmetry
between positively and negatively charged leptons,
\begin{equation}
\label{eq:asym}
   \mathcal{A} = \frac{\ell^+ - \ell^-}{\ell^+ + \ell^-} \,,
\end{equation}
is predicted indeed~\cite{Barr:2004ze}.  [The asymmetry is washed out
somewhat in $pp$ collisions at the LHC where also anti-squarks
${\tilde{q}}^\ast$ are produced; see section~\ref{ssec:pp}].
Fig.~\ref{Fig:asym} shows the parton-level asymmetry for the
${\tilde{q}}_L$ decay chain in SPS1a$'$. In particular for high
invariant masses the dominance of positively charged leptons is
evident. The NLO SQCD radiative corrections included in our decay
analysis have a very small influence on the asymmetry. This is evident
from the lower panel of Fig.~\ref{Fig:inv_mass}; near the kinematic
endpoint where the corrections to the invariant mass distribution of
the positively charged lepton are biggest, the asymmetry is close to
unity and changes are marginal as a result.

\begin{boldmath}
  \subsection{Squark cascade decays in $pp$ collisions at the
    LHC}\label{ssec:pp} The \end{boldmath} more detailed numerical
analysis presented in this final section is also based on the
reference point SPS1a$'$ \cite{Allanach:2002nj} in the mSUGRA
scenario. This point is located in the transition zone from the bulk
to the coannihilation region in the $[M_{1/2},M_0]$ gaugino/scalar
mass plane. It corresponds to the region of maximum probability in the
analysis of electroweak precision observables
\cite{Allanach:2007qk,Ellis:2007fu}, driven by the contribution to the
muon anomalous magnetic moment, $g_\mu - 2$; the reference point is
compatible with the WMAP measurement of the density of cold dark
matter. The parameters of this mSUGRA scenario are defined as follows:
gaugino mass $M_{1/2} = 250$ GeV, scalar mass $M_0 = 70$ GeV, and
trilinear coupling $A_0 = -300$ GeV at the unification scale; moreover
$\tan\beta = 10$ and ${\rm sign}(\mu) = +$.  After evolution down to
the Terascale the squark, gluino, slepton, and neutralino masses that
will be used in the phenomenological analysis are collected in
Table~\ref{ta:SPS_masses}.  While {\tt Suspect}~\cite{Djouadi:2002ze}
has been employed for the computation of the entries in the table, the
{\tt Spheno}~\cite{Porod:2003um} and {\tt
  SOFTSUSY}~\cite{Allanach:2001kg} spectra are quite similar, with
differences at the level of 1\% and less.

Table~\ref{ta:SPS_masses} also presents the branching ratios which are
based on NLO calculations as implemented in {\tt
  Susyhit}~\cite{Djouadi:2006bz} / {\tt
  Sdecay}~\cite{Muhlleitner:2003vg}.  The $L$-squarks decay
predominantly into the charginos $\tilde{\chi}_1^{\pm}$ and the
neutralino ${\tilde{\chi}}_2^0$.  $R$-squarks decay almost exclusively
into the ${\tilde{\chi}}_1^0$ LSP which is bino-like. Hence, almost
all of the squarks decaying to ${\tilde{\chi}}^0_2$ are indeed
$L$-type squarks.  Only a small admixture of $R$-type squarks
diminishes the observed spin correlation and the asymmetry,
Eq.~(\ref{eq:asym}), generated by the ${\tilde{\chi}}_2^0$ decay
chain. More than half ({\it i.e.} 57\%) of the neutralinos
${\tilde{\chi}}_2^0$ decay visibly into charged slepton-lepton-pairs.
Since left-handed sleptons of the first two generations are heavy, the
decays are dominated (${\rm BR}=0.527$) by decays into mixed
$\tilde{\tau}_1$ states \cite{R30A} with a reduced mass eigen-value,
but still a significant fraction decays to $\tilde{e}_R,
\tilde{\mu}_R$ (${\rm BR}=0.024$ for each mode).  If the $e,\mu$
leptonic decays are used as experimental signals, the transverse
momenta of the cascading $\tau \to e,\mu$ final states are strongly
reduced. A large fraction of the $\tau \to e,\mu$ decay events will
thus be lost in experimental cuts, about 90\% for a transverse
momentum cut of 10 GeV.

Since experiments will not distinguish quark from anti-quark jets,
experimentally observable distributions will also receive
contributions from anti-squark decay chains. For anti-squarks, by
$CP$-invariance, the r{$\hat{\rm{o}}$}le of positively and negatively
charged leptons is interchanged and the asymmetry $\mathcal{A}$ is
reduced as a result. At the Tevatron, the equal rate for squark and
anti-squark production even drives $\mathcal{A}$ to zero.  However,
the (non-$CP$-invariant) proton-proton initial state at the LHC
ensures that more squarks than anti-squarks are produced and a
non-zero asymmetry is predicted.

At the LHC, squarks are produced either directly in
quark/anti-quark/gluon collisions, or they are decay products of
gluinos as $M_{\tilde{g}} > M_{\tilde{q}}$ for the SPS1a$'$ reference
point. The production cross sections for both species have been
determined including the next-to-leading order corrections of
Ref.~\cite{Beenakker:1996ch} as implemented in {\tt
  Prospino}~\cite{prospino}.  The NLO calculation of
Ref.~\cite{Beenakker:1996ch} has been performed for mass-degenerate
$L$ and $R$-squarks and mass-degenerate first and second generation
squarks of up and down type. Since the mass differences are in general
small, in particular for the spectrum of the SPS1a$'$ reference point
where $[M^2_L - M^2_R]/[M^2_L + M^2_R] \approx 4\%$, this is a valid
approximation. In our tree-level calculation, to which we apply the
K-factors, the exact kinematics is used. For the parton densities the
CTEQ5 set~\cite{Lai:1999wy} has been adopted and the renormalization
and factorization scales have been set to the average mass of the
produced particles. The production cross sections are displayed for
squarks and anti-squarks of the first two generations in
Table~\ref{ta:SPS_cross}.  The $pp$ initial state at the LHC clearly
favours the direct production of squarks over the direct production of
anti-squarks. Though equal numbers of squarks and anti-squarks are
generated by gluino decays, the ratio of the number of anti-squarks
over squarks is only slightly diminished because the branching ratio
of gluinos into first and second generation $L$-squarks is small.

\begin{table}
\begin{center}
\begin{tabular}{|c|c|cl||c|c|rl|}
\hline
& Mass[GeV] & 
  \multicolumn{2}{c||}{BR($[\tilde{g} \to \,]\, \tilde{q}\to\tilde{\chi}^2_0$)[\%]} & &
  Mass[GeV] &
  \multicolumn{2}{c|}{BR($\tilde{\chi}^0_2 \to \tilde{l}$)[\%]} \\
\hline \hline
$\tilde{u}_L$ & 558 &                & \phantom{0}32.2            & $\tilde{\chi}^0_1$ &  98 &&\\
$\tilde{d}_L$ & 564 &                & \phantom{0}31.6            & $\tilde{\chi}^0_2$ & 183 & 
\mbox{}\hspace{0.2cm}                                                  $\tilde{e}^+_R$:    &  \phantom{0}1.2 \\
$\tilde{u}_R$ & 542 &                & \phantom{00}0.6  &                    &     & $\tilde{\tau}_1^+$: & 26.4 \\ \cline{5-8}
$\tilde{d}_R$ & 542 &                & \phantom{00}0.6  & $\tilde{e}_R$      & 124 &&\\ \cline{1-4}
$\tilde{g}$   & 605 & $\tilde{u}_L/\tilde{d}_L$: & 0.8/0.6 & $\tilde{\tau}_1$   & 108 &&\\ 
              &     & $\tilde{u}_R/\tilde{d}_R$: & 0.03/0.03 &                    &     &&\\
\hline
\end{tabular}
\end{center}
\mycaption{\label{ta:SPS_masses} 
Masses and decay branching ratios most relevant 
for evaluating the SPS1a$'$ squark/gluino cascades.}
\end{table}

\begin{table}
\begin{center}
\begin{tabular}{|c|c|c||rl|c|}
\hline
& $\sigma(pp \to \tilde{q}/\tilde{g})$[pb] &
$\sigma(\tilde{q}/\tilde{g} \to \tilde{\chi}^0_2 )$[pb] &
\multicolumn{2}{c|}{$\sigma({\tilde{l}}_R)$[pb]} &
$\sigma(e,\mu)$[pb] \\\hline\hline
$\tilde{u}_L$/$\tilde{u}_R$     &  14.4/15.4   & 4.6/0.1\phantom{0}  & $\tilde{e}_R+ \tilde{\mu}_R:$   & 0.57 & 0.57\\
$\tilde{d}_L$/$\tilde{d}_R$     &   6.7/7.4    & 2.1/0.05  & $\tilde{\tau}_1:$ & 6.37 & 0.39\\\cline{1-3}
$\tilde{u}_L^*$/$\tilde{u}_R^*$ &   1.7/1.9    & 0.6/0.01 & & &\\
$\tilde{d}_L^*$/$\tilde{d}_R^*$ &   2.0/2.3    & 0.6/0.01 & & &\\\cline{1-3}
$\tilde{c}_L$/$\tilde{c}_R$     &   0.8/0.9    & 0.3/0.01 & & &\\ 
$\tilde{s}_L$/$\tilde{s}_R$     &   1.2/1.4    & 0.4/0.01 & & &\\\cline{1-3}
$\tilde{c}_L^*$/$\tilde{c}_R^*$ &   0.9/1.0    & 0.3/0.01 & & &\\
$\tilde{s}_L^*$/$\tilde{s}_R^*$ &   1.3/1.4    & 0.4/0.01 & & &\\\cline{1-3}
$\tilde{g}$   &   45.1 & 2.6 & & &\\\cline{1-6}
\end{tabular}
\end{center}
\mycaption{\label{ta:SPS_cross} 
  Cross sections for the production of squarks of the first
  two generations and gluino production  in       
  $pp$ collisions at the LHC, and the production cross sections   
  for the neutralino ${\tilde{\chi}}^0_2$ in the cascades. Also
  shown are the cross sections for $R$-sleptons after summing
  up all cascade channels. The resulting cross sections 
  for electron/muon pairs from direct electron/muon production in 
  the decay chain and from tau decays are collected in the 
  right-most column.}
\end{table}

Table~\ref{ta:SPS_cross} also includes the cross sections times
branching ratios $\sigma (\tilde{q}) \times {\rm BR}(\tilde{q} \to
{\chi}^0_2)$ and $\sigma (\tilde{g}) \times {\rm BR}(\tilde{g}
\to \tilde{q} \to {\chi}^0_2)$ for neutralino
${\tilde{\chi}}^0_2$ production from squark decays, and the final
cross sections times branching ratios for the signal $\sigma [e,\mu]$,
{\it i.e.} lepton production from squark cascades, with $\tau$ decays
separated.  With a nominal integrated luminosity of 300 fb$^{-1}$, a
sample of roughly $2 \cdot 10^5$ events is accumulated for the
SPS1a$'$ benchmark for $e,\mu$ in the first two generations and,
before experimental cuts, an initial number of $1.5 \cdot 10^5$ $\tau$ 
events decaying leptonically into either electrons or muons.
Although these calculations are of purely theoretical nature, they
provide nevertheless a solid platform for estimating expectations
before experimental simulations will finally include selection cuts,
efficiencies {\it etc}.

Using these estimates of production cross sections and branching
ratios, the invariant mass distributions for the SPS1a$'$ decay chain
can be calculated. Fig.~\ref{Fig:inv_mass_lplm_full} displays the LO
and NLO $[{\rm jet} \, {\ell}^\pm]$ invariant mass spectrum for
positively and negatively charged leptons as predicted for the LHC.
The distributions are normalized to the signal cross sections and are
shown separately for ${\tilde{\chi}}^0_2$ decays to first/second
generation leptons/sleptons and for $\tau/{\tilde{\tau}}$ decaying
leptonically. $L/R$ mixing of the $\tau$ sleptons and the
$\tilde{\chi}_2^0$ higgsino component exacerbate the analysis of the
$\tau/\tilde{\tau}$ channel considerably; all these effects have been
taken into account in the numerical analysis. However, the leptonic
$\tau$ decay signal is strongly reduced to a level of $10\%$ if a
transverse momentum cut of 10~GeV is applied as indicated by the
dotted lines. [For simplicity, the cut is applied in the squark rest
frame. This approximation is adequate since squarks are produced
predominantly at small velocities.]  The experimentally observable
distributions are clearly distinct despite of the anti-squark
admixture to the squark sample.  As a result of the anti-squark
admixture, the $\ell^+/\ell^-$ asymmetry (\ref{eq:asym}) is slightly
reduced at the LHC compared with Figure~\ref{Fig:asym} for pure
${\tilde{q}}_L$ decays, with marginal NLO corrections as before.
However, the asymmetry is still quite significant for large [${\rm
  jet} \,\ell$] invariant masses as demonstrated in
Refs.~\cite{Barr:2004ze,Smillie:2005ar, Athanasiou:2006ef}, even after
including detector effects. 

\begin{figure}
\begin{center}
\includegraphics[width=17cm]{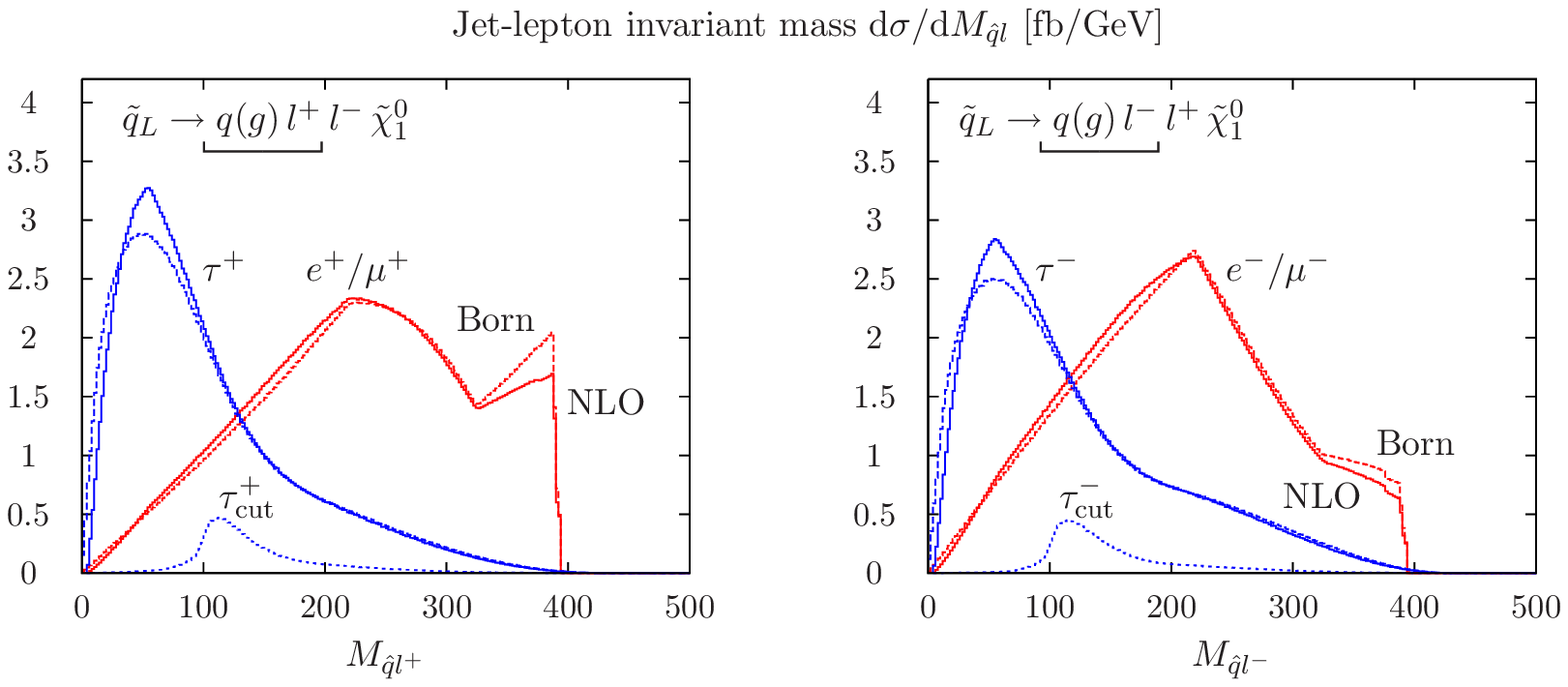}
\end{center}
\mycaption{\label{Fig:inv_mass_lplm_full} Invariant $[{\hat{q}}l]$ mass 
distributions, $y_{\rm c} = 0.01$, for muons and electrons 
produced directly in the neutralino and slepton decay. For muon and 
electron pairs produced from tau decays in the decay chain,
the reduction of the NLO cross sections for a transverse momentum cut
of 10 GeV is indicated by the dotted lines.} 
\end{figure}

\section{Conclusions}\label{sec:conc}
\setcounter{equation}{0}

Edges/thresholds, distributions and correlations of particles in
cascades play an important r\^{o}le in determining properties of
supersymmetric particles, {\it i.e.} masses and spins.
Lepton-lepton and jet-lepton invariant mass distributions are 
particularly useful instruments in this context.

In the present article we have studied how the parton final states in
$\tilde{q} \to q \tilde{\chi}$ decays are affected by super-QCD
corrections at next-to-leading order, in particular by the radiation
of gluons in the decay process.  The corrections modify the energy as
well as the polarization of the particles. By performing an analytic
next-to-leading order analysis we have provided a detailed
understanding of origin and size of these effects.  Moreover, we have
studied particular decay channels and $[{\rm jet} \, {\ell}^\pm]$
invariant mass distributions, exemplified for the SPS1a$'$ benchmark
scenario, in ${\tilde{q}}_L \to q\, \ell^+\ell^- \, {\tilde\chi}^0_1$
decays. We have found significant gluonic corrections to the shape of
the invariant $[{\rm jet} \, \ell^+]$ mass distribution near the
pronounced edge at the kinematic endpoint.

\section*{Acknowledgments}
We gratefully acknowledge communications with G.~Polesello, and
D.J.~Miller, P.~Osland and A.R.~Raklev. We are thankful to A.~Djouadi
for comments on the manuscript. This work was supported in part by the
DFG SFB/TR9 ``Computational Particle Physics", and by the European
Community's Marie-Curie Research Training Network HEPTOOLS under
contract MRTN-CT-2006-035505. PMZ thanks the LPT and the Institut
f\"ur Theoretische Physik E for the warm hospitality extended to him
at the Universit\'{e} Paris-Sud and at the RWTH Aachen.

\end{document}